\documentstyle{article}

\font\tenrm=cmr10

\font\elevenbf=cmbx10 scaled\magstep 1
\font\elevenrm=cmr10 scaled\magstep 1
 1

\renewenvironment{thebibliography}[1]
 { \tenrm
 \baselineskip=10pt
   \begin{list}{\arabic{enumi}.}
    {\usecounter{enumi} \setlength{\parsep}{0pt}
     \setlength{\itemsep}{3pt} \settowidth{\labelwidth}{#1.}
     \sloppy
    }}{\end{list}}

\begin{document}
\begin{center}{\elevenbf
SKYRMIONS FROM SU(3) HARMONIC MAPS AND THEIR QUANTIZATION
\\}
\vglue 0.5cm
{\tenrm V.B. Kopeliovich$\sp{\dagger\star}$, B.E. Stern$\sp{\dagger}$
and W.J.Zakrzewski$\sp{\star}$,\\}
\vglue 0.3cm
{$\sp{\dagger}$Institute for
Nuclear Research
 of the Russian Academy of
Sciences,\\ Moscow 117312, Russia}\\
\vglue 0.3cm
{$\sp{\star}$Department of Mathematical Sciences, University of Durham,\\
Durham DH1 3LE, UK\\}
\vglue 0.5cm
\end{center}
{\rightskip=3pc
\leftskip=3pc
\tenrm\baselineskip=12pt
\noindent
Static properties of $SU(3)$ multiskyrmions with baryon numbers up to $6$
are estimated. The calculations are based on the recently
suggested generalization of the $SU(2)$ rational map ans\"atze applied to the
$SU(3)$ model. Both $SU(2)$ embedded skyrmions and genuine $SU(3)$
solutions are considered and it is shown that although, at the classical 
level, the energy of the embeddings is lower, the quantum corrections can 
alter these conclusions. This correction to the energy of the lowest state, 
depending on the Wess-Zumino term, is presented in the most general case.
 \vglue 0.2cm}
 \vglue 0.2cm
\baselineskip=13pt
\elevenrm
{\bf 1.} Topological soliton models, and the Skyrme model among them
\cite{1}, have recently generated a fair amount of interest because
they may be able to describe various properties of low energy baryons.
So far, most of such studies have involved the $SU(2)$ Skyrme model, whose
solutions were then embedded into the $SU(3)$ model. This is justified
as the solutions of the $SU(2)$ model are also
solutions of $SU(N)$ models. However, there exist solutions
of the $SU(N)$ models which are {\bf not} embeddings of $SU(2)$
fields and it is important to assess their contribution.

As with solutions of the $SU(2)$ Skyrme models, the solutions
of the $SU(3)$ model, with very few exceptions, can only be
determined numerically. Like for the $SU(2)$ case \cite{2} one starts with
a harmonic map ansatz which gives fields ``close" to the genuine
solutions \cite{3,4}. Then one can use these fields as starting configurations
 of various numerical minimization schemes. For the $SU(2)$ fields such an
approach was carried out in \cite{2} where it reproduced the results
of the earlier numerical approaches \cite{5,6} and so showed that the
harmonic map approximations 
are very close to the final fields. The
real solutions have energies only few hundreds of $MeV$ lower
than the harmonic approximations and the baryonic charge and energy
distributions do not look very different {\it etc}.
This has justified the use of harmonic field approximants
in calculating various quantities when estimating quantum corrections
to the classical results.

In this paper we take a look at the nonembedding solutions of the $SU(3)$ model.
First we recall the results of \cite{3} and use them in a $SU(3)$
numerical minimization program to estimate the values of energies
of true solutions. The $SU(3)$ variational minimization program
\cite{7} has been rearranged for this purpose to allow the
consideration of quite general field ans\"atze. This is discussed in the
next section.

In the following section we discuss various quantum corrections and compare
our results with the similar results for the embeddings.
The paper ends with a short section presenting our conclusions.\\

{\bf 2.} The Lagrangian of the $SU(3)$ Skyrme model, in its well known
form,
depends on the parameters $F_{\pi} $ and $e$ and is
given by \cite{8,9}:
$${\cal L} =  \frac{F_\pi^2}{16}Tr l_{\mu} l^{\mu} + {1 \over 32e^2}
Tr [l_\mu,l_\nu]^2 +\frac{F_\pi^2m_\pi^2}{16} Tr(U+U^{\dagger}-2) \eqno (1) $$
Here $U \in SU(3)$ is a unitary matrix describing the chiral (meson) fields, and
$l_\mu=U^{\dagger} \partial _\mu U$. In the model $F_\pi$ should be fixed
at the physical value: $F_\pi$ = $186$ Mev .
The flavour symmetry breaking $(FSB)$ in the Lagrangian will be 
considered below.

The Wess-Zumino term, which should to be added to the action,
is given by
$$ S^{WZ}= \frac{-iN_c}{240 \pi^2}\int_\Omega d^5x\epsilon^{\mu\nu\lambda\rho
\sigma} Tr (\tilde l_\mu \tilde l_\nu \tilde l_\lambda \tilde l_\rho \tilde l_\sigma), \eqno (2) $$
where $\Omega $ is a 5-dimensional region with the 4-dimensional space-time
as its boundary and where $\tilde l_\mu$ is a 5-dimensional
analogue of $l_\mu = U^\dagger \partial_\mu U$.
As is well known, this extra term
does not contribute to the static masses of classical configurations, but it
defines important topological properties of skyrmions \cite{8,9} and plays an
important role in their quantization \cite{10,11}.

In \cite{3} an ansatz was presented which allows us to find approximate
solutions of the $SU(3)$ model. This ansatz involves parametrising the
static field as
$$U(\vec x)\,=\,\exp{\left\{f(r)\left(P(\theta,\varphi)-{1\over 3}\right)\right\}},\eqno(3)$$
where $r,\theta$ and $\varphi$ are polar coordinates,
 $f(r)$ is a radial profile function which has to be determined
numerically and $P(\theta,\varphi)$ is a projector involved in the harmonic
map ansatz. As shown in \cite{3} this projector is given by
$$P\,=\,{FF\sp{\dagger}\over \vert F\vert\sp2},\eqno(4)$$
where $F$ is a 3-component vector, whose entries are polynomials in $z=
tan({\theta\over2})e\sp{i\varphi}.$ The largest degree of the polynomial
gives the baryon number $B$ of the final $SU(3)$ Skyrme field configuration.
For an $SU(2)$ embedding the approach is similar except that this time
we put
$$U(\vec x)=\pmatrix{U_2&\vec 0\cr
\vec 0&1\cr},\eqno(5)$$
where $U_2$ is an $U(2)$ matrix determined in an analogous way as $(3)$ except
that the vector $F$ has only two components.

The fields $F$ in Eq. $(4)$ are so chosen that the configuration $(3)$ has the
smallest energy; this provides us with the harmonic map approximation
to the real minimal energy static solution of the Skyrme model.
We have then taken the expressions for $F$ determined in \cite{3} and used
the corresponding $U$ in $(3)$ as an initial field in our minimization
program.\\
 For $B=2$ $F=(1, \sqrt{2}z, z^2)^T$, for $B=3$ $F=(1/\sqrt{2},1.576z, z^3)^T$,
for $B=4$ $F=(1, 2.72z^2, z^4)^T$,\\
for $B=5$ $F=(4.5z^2,2z^4+1,z^5-2.7 z)^T$ and 
for $B=6$ $F=(kz^3,1-3z^5, z^6+3z)^T$ with $k=7.06$ \cite{3}.

We have performed many minimizations of the field configurations
involving baryon numbers $3,4,5$ and $6$. In all cases the initial $SU(3)$
harmonic map  fields had energies somewhat higher than the corresponding
harmonic map embeddings.

Our $SU(3)$ minimization program, which uses the
parametrization of the $SU(3)$ field in terms of two, mutually
orthogonal, complex unit vectors\footnote{The authors thank W.K. Baskerville
for making them aware of this parametrisation} was then used
to minimize the energy further. The constraints of the orthogonality
and of the vectors being of unit length were replaced by the introduction
of extra positive contributions to the energy (with large coefficients)
- the so called ``penalty terms".

The program itself minimized the energy using a mixture of a finite element
and finite step methods, and we varied the coefficients of the penalty 
terms so as
not to be trapped in a local minimum. In each case the final energy of the field
configuration was lower by a few hundreds of $MeV$ from the energy
of the harmonic approximant. However, due to the smallness of the lattice
the baryon number was also somewhat lower; so our values in the {\bf Table}
were obtained by a linear extrapolation: $M=M+\Delta M$, where $\Delta M
=\alpha*(B-B_{obs})$ and where $B$ is the baryon number and $B_{obs}$ is
its value on the lattice. $\alpha$ was determined by looking at various
minimizations (with different values of the coefficients of the penalty terms).

To check the stability of the program we also performed a series
of minimizations when the initial field was a mixture of the $SU(3)$
harmonic map field and of the embedding. When the mixture was close
to the embedding or to the $SU(3)$ harmonic map the minimization program
took it down to these fields. When the mixture was far from either
of these special fields it evolved to a new configuration of higher
energy than either of the special ones. Although some of these
 new configurations can be numerical artifacts it is clear that the spectrum
of static solutions of the $SU(3)$ model is very complicated, with most
states having energies larger than the energies of the embeddings or the
fields derived by the harmonic map ansatz. Hence our estimates in the 
{\bf Table} have some physical justification.

In the {\bf Table} we present some of our results. It is difficult to assess 
their accuracy; it is probably within $1\%$ for the masses and several $\%$ for
other quantities. In all numerical minimizations we have worked on a small
tetrahedral lattice and hence we had a small ``leakage" of both the energy
and of the baryon number.
\begin{center}

\begin{tabular}{|l|l|l|l|l|l|l|l|l|l|l|l|l|l|l|}
\hline
$B$& $M_{cl}$&$M_{cl}^{SU_2}$&$\Delta E_{M.t.}$ &$C_S$ & $\Theta_{1,2}$ & $\Theta_3$ & 
$\Theta_4$&$\Theta_5$& $\Theta_{6,7}$&$\Theta_8$ &$\Theta_{38}$&
$WZ_3$&$WZ_8$&$\Delta E$ \\
\hline
1 & $1.70$&$1.70$&$46$&---& $5.56$ &$5.56$ &$2.04$ &$2.04$ &$2.04$
&$-$&$-$&-&-&$368$
\\
\hline
2 &$3.38$&$3.26$
&$89$&$0.33$&$5.70$&$6.40$&$7.10$&$7.10$&$5.70$&$5.0$&$-1.2$&$0.00$&$0.00$&
$0.0 $\\
\hline
3&$4.85$&$4.80$&
$110$&$0.33$&$7.15$&$7.90$&$8.60$&$8.60$&$9.20$&$5.1 $&$-1.9$&$-0.7$&
$-0.15$&$35$\\
\hline
4&$6.48$&$6.20$&$174$&$0.31$&$12.2$&$9.80$&$11.6$&$11.6$&$12.1$&$6.2$&$-3.0$&
$-1.0$&$0.6$&$80$ \\
\hline
5&$7.90$&$7.78$&$211$&$0.36
$&$11.5$&$11.6$&$15.4$&$12.1$&$12.7$&$15.2$&$-0.1$&$0.2$&$-0.8$
&$23$\\
\hline
6&$9.38$&$9.24$&$251$&$0.33 $&$15.4$&$14.8$&$15.5$&$15.5$&$14.2$&$16.3$&$-1.2$&
$-0.01$ &$-0.03$&$0.0$\\
\hline
\end{tabular}
\end{center}
\vspace{1mm}
\baselineskip=11pt
{\bf Table.} {\tenrm The values of the masses $M_{cl}$ in $GeV$, the mass term
$\Delta E_{M.t.}$ (in $MeV$), the strangeness content $C_S$ and the moments of
 inertia (in $GeV^{-1}$) for the $SU(3)$ projector configurations.
The quantities for $B=1$ hedgehog and the masses of
$SU(2)$ embeddings $M_{cl}^{SU_2}$, in $GeV$ are given for comparison.
The components of the Wess-Zumino term $WZ_3$ and $WZ_8$ which are 
different from zero for the $SU(3)$ projector ansatz are also shown.
$\Delta E$ in the last column is the quantum correction (in $MeV$) due to 
zero modes for the lowest  energy state.
The parameters of the model are $F_\pi=186 \,MeV,\; e=4.12$.}\\

\baselineskip=13pt
As can be seen from this {\bf Table}, the moments of inertia are
pairwise equal, except for the case $B=5$ when $\Theta_4 \ne \Theta_5$.
These equalities are a consequence of the symmetry properties of
our multiskyrmion configurations.

It should be noted that the energy of the $B=4$ configuration is
very close to that of the $B=4$ toroidal soliton obtained in $1988$
\cite{12} within the axially symmetrical generalization of the ansatz by
Balachandran et al. \cite{13}. Within the accuracy of our calculations we
cannot be certain which configuration has the lowest energy in the chirally
symmetrical limit.
For $B=6$ the torus-like $SO(3)$ configuration has energy considerably
higher than the energy of the $SU(3)$ projector ansatz presented in the {\bf 
Table}.

Let us add here that the configuration of a double-torus form
was considered also within the $SU(2)$-version of the model several years
ago \cite{14}. It has the energy somewhat higher
than the energy of the known minimal energy configuration.\\

{\bf 3.} Following \cite{10,11,15} we consider the contribution of the
Wess-Zumino 
$(WZ)$ term which defines the quantum numbers of the system in the quantization
procedure. Its expression is given by $(2)$ above. As usually, we
introduce the time-dependent collective coordinates for the quantization
of zero modes according to the relation:
$ U(\vec{r},t)=A(t)U_0(\vec{r})A^{\dagger} (t)$. Next we perform some
integration by parts and rewrite the expression for the WZ-term as:
$$L^{WZ}=\frac{-iN_c}{48\pi^2} \epsilon_{\alpha\beta\gamma}\int
Tr A^{\dagger}\dot{A} (R_{\alpha}R_{\beta}R_{\gamma} +
L_{\alpha}L_{\beta}L_{\gamma}) d^3x , \eqno (6) $$
where $L_{\alpha}=U_0^{\dagger}\partial_{\alpha}U_0=iL_{k,\alpha}\lambda_k$ 
and $R_\alpha=\partial_{\alpha}U_0U_0^{\dagger}=U_0 L_{\alpha} U_0^{\dagger}$, 
or 
$$L^{WZ}=\frac{N_c}{24\pi^2}\int \sum_{k=1}^{k=8} \omega_k WZ_k d^3x=
\sum_{k=1}^{k=8} \omega_kL^{WZ}_k , \eqno (7) $$
with the angular velocities of rotation in the configuration space defined
in the usual way: \\
$A^{\dagger}\dot{A}=-{i \over 2} \omega_k\lambda_k$.
Summation over repeated indices is assumed here and below.
The functions $WZ_k$ can be expressed through the chiral derivatives
$\vec{L}_k$:
$$ WZ_i= WZ_i^R+WZ_i^L=
(R_{ik}(U_0)+\delta_{ik})WZ_k^L, \eqno (8a)  $$
$i,k=1,...8$, and are given by \cite{15}
$$WZ_1^L=-(L_1,L_4L_5+L_6L_7)-(L_2L_3L_8)/\sqrt{3}-
2(L_8,L_4L_7-L_5L_6)/\sqrt{3} $$
$$WZ_2^L=-(L_2,L_4L_5+L_6L_7)-(L_3L_1L_8)/\sqrt{3}-
2(L_8,L_4L_6+L_5L_7)/\sqrt{3} $$
$$WZ_3^L=-(L_3,L_4L_5+L_6L_7)-(L_1L_2L_8)/\sqrt{3}-
2(L_8,L_4L_5-L_6L_7)/\sqrt{3} $$
$$WZ_4^L=-(L_4,L_1L_2-L_6L_7)-(L_3L_5L_8)/\sqrt{3}+
2(\tilde{L}_8,L_1L_7+L_2L_6)/\sqrt{3} $$
$$WZ_5^L=-(L_5,L_1L_2-L_6L_7)+(L_3L_4L_8)/\sqrt{3}-
2(\tilde{L}_8,L_1L_6-L_2L_7)/\sqrt{3} $$
$$WZ_6^L=(L_6,L_1L_2+L_4L_5)+(L_3L_7L_8)/\sqrt{3}-
2(\tilde{\tilde{L}}_8,L_1L_5-L_2L_4)/\sqrt{3} $$
$$WZ_7^L=(L_7,L_1L_2+L_4L_5)-(L_3L_6L_8)/\sqrt{3}+
2(\tilde{\tilde{L}}_8,L_1L_4+L_2L_5)/\sqrt{3} $$
$$WZ_8^L= -\sqrt{3}(L_1L_2L_3)+(L_8L_4L_5)+(L_8L_6L_7), \eqno(9) $$
where $(L_1,L_2L_3)$ denotes the mixed product of vectors
$\vec{L}_1$, $\vec{L}_2$, $\vec{L}_3$,
{\it ie} $(L_1,L_2L_3)=(\vec {L}_1\cdot \vec {L}_2\wedge\vec {L}_3)$  and
$\tilde{L}_3=(L_3+\sqrt{3}L_8)/2$,
$\tilde{L}_8=(\sqrt{3}L_3-L_8)/2$, $\tilde{\tilde{L}}_3=(-L_3+\sqrt{3}L_8)
/2$, $\tilde{\tilde{L}}_8=(\sqrt{3}L_3+L_8)/2$ are the third and eighth
          components
of the chiral derivatives in the $(u,s)$ and $(d,s)$ $SU(2)$-subgroups.
$R_{ik}(U_0)={1 \over 2}Tr\lambda_iU_0\lambda_kU^{\dagger}_0$ is a real
          orthogonal matrix, and
$WZ^R_i$ are defined by the expressions $(9)$ with the substitution $\vec{L}_k
          \rightarrow \vec{R}_k$. Relations similar to $(9)$ can be
          obtained for $\widetilde{WZ}_3$ and $\widetilde{WZ}_8$; they are
          analogs
          of $WZ_3$ and $WZ_8$ for the $(u,s)$ or $(d,s)$ $SU(2)$ subgroups,
      thus clarifying the symmetry of the $WZ$-term in the different $SU(2)$
        subgroups of $SU(3)$.

The baryon number of the $SU(3)$ skyrmions can be written also in terms of
$\vec{L_i}$ in a form where its symmetry in the different $SU(2)$ subgroups
of $SU(3)$ is more explicit:
$$ B=-\frac{1}{2\pi^2} \int \biggl(({L_1},{L_2}{L_3}) +
({L_4},{L_5}\tilde{{L_3}}) +
({L_6},{L_7}\tilde{\tilde{{L_3}}})+\frac{1}{2}
[({L_1},{L_4}{L_7}-{L_5}{L_6})+
({L_2},{L_4}{L_6}+{L_5}{L_7})]\biggr) d^3r. \eqno(10) $$
The contributions of the three $SU(2)$ subgroups enter the baryon number
          on an equal
footing. In addition, mixed terms corresponding to the contribution of
the chiral fields from different subgroups are also present.

The results of calculating the $WZ$-term according to
$(9)$ depend on the orientation of the soliton in the $SU(3)$
configuration space.
The Guadagnini's quantization condition \cite{10} was generalized in 
\cite{15} for configurations of the ``molecular" type to
$$ Y^{min}_R = {2 \over \sqrt{3}} {\partial L^{WZ}\over \partial\omega_8 }
\simeq {N_c B (1 - 3C_S)\over 3},      \eqno(11)  $$
where $Y_R$ is the so-called right hypercharge characterizing the
$SU(3)$ irrep under consideration,
and the scalar strangeness content $C_S$ is defined in terms of the
real parts of the diagonal matrix elements of the matrix $U$:
$$C_S=\frac{<1-Re U_{33}>}{<3-Re(U_{11}+U_{22}+U_{33})>}, \eqno(12) $$
where $<>$ denotes the averaging or integration over the whole 3-dimensional 
space. 
When solitons are located in the $(u,d)$ $SU(2)$ subgroup of $SU(3)$
only $\vec{L}_1$, $\vec{L}_2$ and $\vec{L}_3$ are different from zero, $C_S=0$, $WZ_8^R$ and
$WZ_8^L$ are both proportional to the $B$-number density, and the well known
quantization condition \cite{10} takes place
$$ Y_R=  N_c B/ 3.  \eqno(13) $$
The interpolation $(11)$ does not work so well for configurations we
consider here.

The expression for the rotation energy density of the system depending
on the angular velocities of rotations in the $SU(3)$ collective
coordinate space defined in Section 2 can be written in the following
compact form \cite{15}:
$$L_{rot}=\frac{F_{\pi}^2}{32}\bigl(\tilde{\omega}_1^2+\tilde{\omega}_2^2+
..+\tilde{\omega}_8^2 \bigr)+ $$
$$+{1 \over 16e^2} \Biggl\{
(\vec{s}_{12}+\vec{s}_{45})^2 +
(\vec{s}_{45} +\vec{s}_{67})^2 +
(\vec{s}_{67} -\vec{s}_{12})^2
 + {1 \over 2} \biggl(
(2\vec{s}_{13}-\vec{s}_{46} -\vec{s}_{57})^2 +
(2\vec{s}_{23}+\vec{s}_{47} -\vec{s}_{56})^2 + $$
$$+(2\tilde{\vec{s}_{34}}+\vec{s}_{16} -\vec{s}_{27})^2 +
  (2\tilde{\vec{s}_{35}}+\vec{s}_{17} +\vec{s}_{26})^2 +
(2\tilde{\vec{s}_{36}} +\vec{s}_{14} +\vec{s}_{25})^2 +
(2\tilde{\vec{s}_{37}} +\vec{s}_{15} -\vec{s}_{24})^2
\biggr) \Biggr\}, \eqno (14a) $$
or
$$L_{rot}= V_{ik}\tilde{\omega}_i \tilde{\omega}_k/2=
V_{ik}g_{li}g_{mk}\omega_l\omega_m/2 \eqno (14b) $$
Here $\vec{s}_{ik}=\tilde{\omega}_i \vec{L}_k - \tilde{\omega}_k
\vec{L}_i $, $i,k=1,2...8$ are the $SU(3)$ indices,
and $\tilde{\vec{s}_{34}}=(\vec{s}_{34}+\sqrt{3}\vec{s}_{84})/2$,
$\tilde{\vec{s}_{35}}=(\vec{s}_{35}+\sqrt{3}\vec{s}_{85})/2$,
$\tilde{\vec{s}_{36}}=(-\vec{s}_{36}+\sqrt{3}\vec{s}_{86})/2$,
$\tilde{\vec{s}_{37}}=(-\vec{s}_{37}+\sqrt{3}\vec{s}_{87})/2$, {\it ie}
similarly  to the definitions of $\tilde{L}_3$ and $\tilde{L}_8$.
$g_{li}$ are given in $(15)$.
To get $(14)$ we have used the identity: $\vec{s}_{ab}\vec{s}_{cd}-
\vec{s}_{ad}\vec{s}_{cb} = \vec{s}_{ac}\vec{s}_{bd}$. The formula $(14)$
possesses remarkable symmetry relative to the different $SU(2)$
          subgroups of $SU(3)$.
The functions $L_8$ or $\tilde{L}_8$ do not enter $(14)$ nor the expression
$(10)$ for the baryon number density.
The functions $\tilde{\omega}_i$ are connected to the body-fixed
angular velocities of $SU(3)$ rotations by the transformations
$ \hat{\tilde{\omega}}=U_0^{\dagger}\hat{\omega}U_0 - \hat{\omega}, $
or, equivalently
$$ \tilde{\omega}_i=(R_{ik}(U_0^{\dagger})-\delta_{ik})\omega_{k}=
g_{ki}\omega_k. \eqno(15)$$
Here $R_{ik}(V^{\dagger})=R_{ki}(V)$ is a real orthogonal
matrix, $i,k=1,...8$, and $\tilde{\omega}_i^2 = 2(\omega_i^2 - R_{kl}(U_0)
\omega_k\omega_l)$. 

The expression for the static energy can be obtained from $(14)$ by means of
the substitution $\tilde{\omega}_i \rightarrow 2L_i$ and
$\vec{s}_{ik} \rightarrow 2\vec{n_{ik}}$, \cite{15}
with $\vec{n}_{ik}$ being the cross
product of $\vec{L}_i$ an $\vec{L}_k$ {\it ie}
$\vec{n}_{ik}=\vec{L}_i\wedge\vec{L}_k$. From $(14)$ we have then the 
inequality \cite{15}
$$ E_{stat}- M.t. \geq 3\pi^2 B (F_{\pi}/e),   \eqno(16) $$
which was obtained first by Skyrme \cite{1} for the $SU(2)$ model.

Eight diagonal moments of inertia and $28$ off-diagonal ones define
          the rotation
energy, a quadratic form in $\omega_i\omega_k$ as follows from $(14)$ and 
$(15)$. The analytical expressions for the moments of inertia are too lengthy
to be reproduced here. Fortunately, it is possible to perform calculations
without explicit analytical formulas, by substituting $(15)$ into $(14)$.

For configurations generated by $SU(3)$ projectors the Lagrangian of the
system can be written in terms of  angular velocities
of rotation and moments of inertia in the form (in the body-fixed system):
$$ L_{rot}=\frac{\Theta_1}{2}(\omega_1^2+\omega_2^2)+\frac{\Theta_3}{2}
\omega_3^2
+\frac{\Theta_4}{2}\omega_4^2+\frac{\Theta_5}{2}\omega_5^2+\frac{\Theta_6}{2}
(\omega_6^2+ \omega_7^2)+ \frac{\Theta_8}{2}\omega_8^2 +
\Theta_{38}\omega_3\omega_8+ WZ_3\omega_3 +WZ_8\omega_8.   \eqno(17) $$

After the standard quantization procedure the Hamiltonian of the system,\\
$H=\omega_i \partial L/\partial\omega_i - L$, is a bilinear function of
the generators $R_i$ of $SU(3)$ rotations:
$$ H = \frac{R_1^2+R_2^2}{2\Theta_1} + \Theta_8\frac{(R_3-WZ_3)^2}{2 D_{38}}+
\frac{R_4^2}{2\Theta_4}+ \frac{R_5^2}{2\Theta_5} +
\frac{R_6^2+R_7^2}{2\Theta_6}+ $$
$$ +\Theta_3 \frac{(R_8-WZ_8)^2}{2 D_{38}} 
-\frac{\Theta_{38}}{D_{38}}(R_3-WZ_3)(R_8-WZ_8), \eqno(18) $$
where $D_{38} = \Theta_3 \Theta_8 - \Theta_{38}^2 $. 
For the states belonging to a definite $SU(3)$ irrep the
rotation energy can be written in terms of the second order Casimir
operators of the $SU(2)$ and $SU(3)$ groups:
$$E_{rot} = 
\frac{N(N+1)-R_3^2}{2 \Theta_1}+\frac{U(U+1)-R_{3,us}^2}{2 \bar{\Theta}_4}
 +\frac{V(V+1)-R_{3,ds}^2}{2 \Theta_6}+$$
$$+\frac{\Theta_8(R_3-WZ_3)^2}{2D_{38}}+ \frac{\Theta_3(R_8 - WZ_8)^2}
{2 D_{38}}-\frac{\Theta_{38}}{D_{38}}(R_3-WZ_3)(R_8-WZ_8)+\frac{(R_4^2-R_5^2)
(\Theta_5-\Theta_4)}{4\Theta_4\Theta_5}
  \eqno (19)  $$
with $\bar{\Theta_4}=2\Theta_4\Theta_5/(\Theta_4+\Theta_5)$.
 $$R_{3,us}=R_{3}/2+\sqrt{3}R_{8}/2  ,\; 
R_{8,us}= \sqrt{3}R_{3}/2 -R_{8}/2, $$
$$R_{3,ds}=-R_{3}/2+\sqrt{3}R_{8}/2,\; 
R_{8,ds}= \sqrt{3}R_{3}/2 +R_{8}/2, \eqno (20) $$ 
with $R_3=R_{3,ud}$, $R_8=R_{8,ud}$;
 $\,R_{3,ab},\, R_{8,ab}$ denote the $3\sp{rd}$ and $8\sp{th}$ generators of the 
$(a,b)$ $SU(2)$ subgroup of $SU(3)$.

$N,\; U$ and $V$ are the values of the right isospin in the $(u,d) \; SU(2)$ 
subgroup, and so called $U$-spin and $V$-spin. They are connected with the
 second order Casimir operator of the $SU(3)$ group 
$C_2(SU_3)={1 \over 3}(p^2+q^2+pq)+p+q$; 
$p,q$ being the numbers of the upper and low indices
in the tensor describing the $SU(3)$ irrep $(p,q)$ via the symmetric 
relation $C_2(SU_3)=N(N+1)+ U(U+1)+ V(V+1)-(R_3^2+R_8^2)/2$. 
The hypercharge $Y=2 R_8/\sqrt{3}$.
The terms linear in the angular velocities present in the Lagrangian due to 
the Wess-Zumino term cancel in the Hamiltonian.

For a state with the lowest energy which belongs to the $SU(3)$ singlet with
$(p,q)=0$ the quantum correction simplifies to:
$$ \Delta E = \frac{1}{2D_{38}} ( \Theta_8 WZ_3^2 + \Theta_3 WZ_8^2-
 2 \Theta_{38} WZ_3 WZ_8 ). \eqno (21a) $$
The quantities $D_{38}$, $\Theta_3+\Theta_8$, $WZ_3^2+WZ_8^2$ do not depend 
on the choice of the particular subgroup of $SU(3)$ group, therefore 
$Eq. (21)$ also is invariant, as one expects from general arguments, since
$$\Theta_8 WZ_3^2+\Theta_3 WZ_8^2 -2\Theta_{38}WZ_3WZ_8 = (\Theta_8+\Theta_3)
(WZ_3^2+WZ_8^2)-\Theta_3 WZ_3^2-\Theta_8 WZ_8^2-2\Theta_{38}WZ_3WZ_8. $$
The generalization of $(21a)$ to an  arbitrary case is
$$ \Delta E^{min} = \Theta_{ij}^{-1} WZ_i WZ_j/2 , \eqno (21b) $$
where $\Theta_{ij}^{-1}$ is the matrix inverse to the tensor of inertia
$\Theta_{ij}$. Eq. $(21b)$ is valid for all cases except the degenerate
case when $Det \Theta_{ij} =0$. Note, that the case of the $SU(2)$ embedding
just corresponds to the degenerate case, with $\Theta_8=0$, and this leads to
the rigorous quantization condition \cite{10}.

As  can be seen from
the {\bf Table}, for the configurations we consider here and for our choice
of the subgroups $\Theta_{38}^2 \ll 
\Theta_3 \Theta_8 $, so a small correction due to $\Theta_{38}$ can be 
neglected, and $\Delta E \simeq WZ_3^2/(2\Theta_3)+WZ_8^2/(2\Theta_8)$. 
Since the moments of inertia are of the order of magnitude $\sim 10 \,
GeV^{-1}$ and both $WZ_3$ and $WZ_8$ are not greater $\sim 1$,
the quantum correction for the lowest states does not exceed
several tens of $MeV$: see the {\bf Table}.

The quantum correction from the $SU(3)$ zero modes which should be added to
the classical energy of the soliton located originally in the $(u,d)$
$SU(2)$ subgroup equals to $\Delta E = 3B/4 \Theta_F \sim 365 \, MeV$ \cite{16}.

Note, that for the $SO(3)$ solitons considered in \cite{13,11} the $WZ$-term
is equal to zero, and for this reason the lowest quantized state was
an $SU(3)$ singlet without any quantum correction to its mass.\\

{\bf 4.} The $FSB$ part of the mass terms in the Lagrangian density which
defines, in
particular, the mass splittings inside $SU(3)$ multiplets is defined, as 
usual, by
$$L_{FSB}= \frac{F_D^2m_D^2-F_\pi^2m_\pi^2}{24}Tr(1-\sqrt{3}\lambda_8)(U+U^{\dagger}-2)+
 \frac{F_D^2-F_\pi^2}{48} Tr(1-\sqrt{3}\lambda_8)(Ul_\mu l^\mu +
l_\mu l^\mu U^{\dagger}) \eqno (22) $$

Here $m_D$ is the mass of $K, \, D$ or
$B$ meson. The ratios $F_D/ F_\pi$ are known to be $1.22$ and $1.7 \pm
0.2$ for, respectively, kaons and $D$-mesons.
The $L_{FSB}$ given by $(22)$ is sufficient to describe the mass
splittings of the octet and decuplet of baryons
within the collective coordinate quantization approach \cite{17}.

The contribution to the classical mass of the solitons from the $SU(3)$
projector ansatz which comes from $FSB$ part of the Lagrangian $(22)$,
considered as a perturbation, is not small even for strangeness:
$$ \Delta M = 2 C_S \Delta E_{M.t} \biggl(\frac{F_K^2 m_K^2}{F_\pi^2 m_\pi^2}
 -1\biggr) \eqno (23) $$
where $\Delta E_{M.t}$ is the $FS$ mass term contribution shown in the {\bf Table}.
Numerically, it equals  $1.13 \, GeV$ for $B=2$ and $3.2 \, GeV$ for
$B=6$. Thus it makes sense to include the $FSB$ part of the mass term into 
the minimization procedure.

We do not consider here the quantum corrections due to the rotations in the 
ordinary space because they are explicitily zero for the lowest states with
even $B$ ($J=0$), or small \cite{18}.

Our investigations have shown that the space of local minima for the
$SU(3)$ skyrmions has a rather complicated form. In addition to the known
local minima - the $SU(2)$ embeddings, $SO(3)$ solitons and skyrmion
molecules - there are also other local minima best approximated by the 
$SU(3)$ projector ansaetze \cite{3,4}.

It is possible to quantize these configuratons and to estimate
the spectrum of states by means of the collective coordinate quantization
procedure. The total sum of the classical mass and of the zero modes quantum
corrections can be the smallest for some baryon numbers.
However, to draw the final conclusions one needs to calculate the Casimir
energies of solitons which are essentially different for solitons of
different form and size. This is a very complicated problem solved 
approximately \cite{19} -\cite{21} only for the $B=1$ hedgehog configurations.
For the case of $SU(2)$ embeddings it was, however, possible to draw
conclusions of physical relevance under the natural assumption that the
unknown Casimir energy, or loop corrections, cancel in the difference
of energies of states with different flavours, and that the states with $u,d$
flavours can be identified with ordinary nuclei \cite{22}. But 
this assumption may be incorrect for the states generated by the $SU(3)$ 
projector ansatz.

We are thankful to W.K. Baskerville, T. Ioannidou, B. Piette and
P.M. Sutcliffe for their interest and helpful discussions.

This work has been supported by the UK PPARC grant: PPA/V/S/1999/00004.
 \\
\vglue 0.2cm
{\elevenbf\noindent References}
\vglue 0.1cm


\begin{thebibliography}{40}
\bibitem{1} T.H.R. Skyrme, Proc. Roy. Soc., London, A260,127 (1961);
Nucl.Phys. 31, 556 (1962)
\bibitem{2} C. Houghton, N. Manton, P.M. Sutcliffe, Nucl. Phys. B510, 507 (1998)
\bibitem{3} T. Ioannidou, B. Piette, W.J. Zakrzewski, J. Math. Phys. 40, 
6353 (1999); Proc. of Internat. Conference "Mathematical methods in modern
physics", Tbilisi, 1998, pp. 91-123, hep-th/9811071
\bibitem{4} T. Ioannidou, B. Piette, W.J. Zakrzewski, 
J. Math. Phys. 40, 6223 (1999); hep-th/9904160
\bibitem{5} E. Braaten, S. Townsend, L. Carson, Phys. Lett. B235,147 (1990)
\bibitem{6} R.A. Battye, P.M. Sutcliffe, Phys. Lett. B391, 150 (1997);
 Phys.Rev.Lett. 79, 363 (1997)
\bibitem{7} V.B. Kopeliovich, B. Schwesinger, B.E. Stern, JETP Lett. 62,
 185 (1995)
\bibitem{8} E. Witten, Nucl. Phys. B223, 422; 433 (1983)
\bibitem{9} G.S. Adkins, C.R. Nappi, E. Witten, Nucl. Phys. B228,552 (1983);
G.S. Adkins, C.R. Nappi, Nucl. Phys. B233, 109 (1984)
\bibitem{10} G. Guadagnini, Nucl. Phys. B236, 35 (1984)
\bibitem{11} R.L. Jaffe, C.L. Korpa, Nucl. Phys. B258, 468 (1985);
 A.P. Balachandran et al, Nucl. Phys. B256, 525 (1985)
\bibitem{12} A.I. Issinsky, V.B. Kopeliovich, B.E. Stern, Sov. J. Nucl. Phys.
48, 133 (1988)
\bibitem{13} A.P. Balachandran et al, Phys. Rev. Lett. 52, 887 (1984)
\bibitem{14} E. Sorace, M. Tarlini, Phys. Lett. 232B, 154 (1989)
\bibitem{15} V.B. Kopeliovich, JETP 85, 1060 (1997); 
 Nucl. Phys. A639, 75c (1998); Phys. Atom. Nucl.56, 1084 (1993)
\bibitem{16} V.B. Kopeliovich, Phys. Lett. B259, 234 (1991);
 V.B. Kopeliovich, B. Schwesinger, B.E. Stern, Nucl. Phys. A549,485 (1992)
\bibitem{17} B. Schwesinger, H. Weigel, Phys. Lett. B267, 438 (1991);
H. Weigel, Int. J. Mod. Phys. A11, 2419 (1996)
\bibitem{18} P. Irwin, hep-th/9804142; J.P. Garrahan, M. Schvellinger,
N.N. Scoccola, hep-ph/9906432
\bibitem{19} B. Moussalam, Ann. of Phys. (N.Y.) 225, 264 (1993);
F. Meier, H. Walliser, Phys. Rept. 289, 383 (1997)
\bibitem{20} H. Walliser, Phys. Lett. B432, 15 (1998)
\bibitem{21} N. Scoccola, H. Walliser, Phys. Rev. D58, 094037 (1998)
\bibitem{22} V.B. Kopeliovich, W.J. Zakrzewski, JETP Lett. 69, 721 (1999)
[Pis'ma v ZhETF 69, 675 (1999)]; hep-ph/9909365
\end{thebibliography}
\end{document}